# Energy Localization Efficiency in TATB Pore Collapse Mechanisms

Brenden W. Hamilton*, Timothy C. Germann

Theoretical Division, Los Alamos National Laboratory, Los Alamos, New Mexico 87545, USA

* brenden@lanl.gov

## Abstract

Atomistic and continuum scale modeling efforts have shown that shock induced collapse of porosity can occur via a wide range of mechanisms dependent on pore morphology, shockwave pressure, and material properties. The mechanisms that occur under weaker shocks tend to be more efficient at localizing thermal energy, but do not result in high, absolute temperatures or spatially large localizations compared to mechanisms found under strong shock conditions. However, the energetic material TATB undergoes a wide range of collapse mechanisms that are not typical of similar materials, leaving the collapse mechanisms and the resultant energy localization from collapse, i.e., hotspots, relatively uncharacterized. Therefore, we present pore collapse simulations of cylindrical pores in TATB for a wide range of pore sizes and shock strengths that trigger viscoplastic collapses that occur almost entirely perpendicular to the shock direction for weak shocks and hydrodynamic-like collapses for strong shocks that do not break the strong hydrogen bonds of the TATB basal planes. The resulting hotspot temperature fields from these mechanisms follow trends that differ considerably from other energetic materials; hence we compare them under normalized temperature values to assess the relative efficiency of each mechanism to localize energy. The local intra-molecular strain energy of the hotspots is also assessed to better understand the physical mechanisms behind the phenomena that leads to a latent potential energy.

# 1. Introduction

The shock initiation of energetic materials is typically governed by the formation of hotspots (excess energy localizations due to shock compression), and is therefore a microstructure dependent phenomena[1]. Hotspot formation mechanisms range from the collapse of porosity, to shear band formation, interfacial friction, and cracking[2]. Shock desensitization experiments have shown pore collapse to be the key mechanism[3,4], where the pressure-volume work conducted during collapse is crucial to reaching extreme temperatures[5]. Despite their significance, experimental techniques are just reaching the necessary temporal and spatial resolutions to resolve individual hotspots and measure their temperatures[6–9]. Hence, theory and simulation techniques have so far led the way in understanding the mechanisms of hotspot formation[10–18] and their resultant chemical reactions[19–25].

Despite the fact that the formation of a hotspot via the collapses of a cylindrical pore is one of the most widely studied problems in shock initiation of energetic materials[11,13,33–35,15,26–32] there are still significant unknowns into the overall factors that lead to larger/hotter hotspots in this idealized pore collapse system. Simulations of shock collapse of 1D planar gaps showed that the strength of the shock and the size of the gap were the controlling factors, with the key contributor being how low of a density the material could expand to in the void, prior to recompression, maximizing the pressure-volume (P-V) work during hotspot formation[5]. When moving to 2D, shock focusing effects and pore geometry come into play. Typically, in cylindrical pores, weak shocks lead to a viscoplastic type collapse[36], while stronger shocks lead to a hydrodynamic mechanism[37]. Both atomistic and continuum modeling efforts in HMX, which used normalized temperature values based on shock strength, showed that the viscoplastic mechanism is more efficient than the hydrodynamic[35,38]. However, since the viscoplastic mechanism is only activated for weak shocks and small pore sizes[26], it does not always result in hotspots relevant to initiation and ignition of energetic materials as the explicit temperature is not sufficient for prompt chemistry.

A wide variety of material properties and local structure can affect the exact collapse process that occurs. High aspect ratio defects can lead to molecule ejecta that accelerate out ahead of the hydrodynamic collapse and maximizes the work done during recompression[38]. Amorphous materials also can lead to significantly different collapse mechanisms and hotspots due to altered sound speed and material strength that effects both the viscoplastic and hydrodynamic collapse routes[39]. Material anisotropy has been shown to significantly change the collapse mechanisms and hotspot formation based on the orientation of the shocked material[13,40], as well as the long timescale evolution of the hotspot[41]. The melting point and viscosity of the material under shock loading plays a significant role in the collapse and hotspot formation[33]. Atomistic scale studies focusing on more complex microstructures with non-trivial void shapes are just beginning to occur[42–45].

Each of these varying collapse mechanisms leads to a different spatial distribution of temperature. While the mapping between collapse mechanism and hotspot temperature field are not well understood, even less understood is the resultant intra-molecular strain energy that results due to large perturbations in molecular shape[46]. The spatial location of peak temperatures and peak intra-molecular strain energies are not the same, and many molecules with the same temperature can possess varying levels of distortion and strain[40]. These resultant molecular strains can accelerate chemistry above what is expected from temperature alone and can vary the reaction paths[47]. Recreating these deformed molecules using steered molecular dynamics has shown that straining different degrees of freedom in the molecule leads to a different level of mechanochemical acceleration and different, alternate first step reaction paths[48]. Hence, the overall reactivity of hotspots stems from its temperature and pressure, the intra-molecular strain that drives mechanochemistry, and the hotspot shape that controls thermal transport away from the reaction, making small changes in hotspot formation critical in understanding chemical initiation.

Specifically, hotspot formation in 1,3,5-trinitro-2,4,6-triaminobenzene (TATB), the energetic material studied here, is complex due to its high anisotropy and its strong hydrogen bonding network along the basal planes. These strongly connected planes of molecules undergo complex defect formation such as buckling, twinning, and non-basal glide[49]. The defects, especially the plane buckling, can lead to large volumetric changes while still keeping the hydrogen bonding network intact. TATB, under low shock strengths, is known to undergo a complex viscoplastic response[13,40]. Instead of the typical, cavitation like event, where pressure surrounds the pore and it collapses radially, the high anisotropy leads to a collapse of the material lateral to the shockwave, where material is effectively extruded from the sides of the pore. This has been shown to occur for multiple TATB orientations, both parallel and perpendicular to the basal planes[13,40], as well as in CL-20[50] and some, but not all, orientations of HMX[35]. This mechanism will be referred to here as the 'extrusion mechanism', shown in the top row of Figure 2, where the more common viscoplastic mechanism will be referred to as the 'radial mechanism'. Hence, in this work, we conduct shock induced pore collapse simulations on cylindrical pores of varying sizes in TATB for multiple shock strengths. The crystal orientation is chosen to study these effects on the complex, extrusion like viscoplastic collapse and how the strength of the basal planes effects the hydrodynamic mechanisms. These trends are utilized to understand the differences in the resulting hotspots.

## 2. Methods

All-atom molecular dynamics simulations were conducted using the LAMMPS package[51,52]. Interatomic interactions were handled with the nonreactive, nonpolarizable force field for TATB[53]. The original force field has since been updated to include tailored harmonic bond stretch and angle bends for fully flexible molecules[54] and an intramolecular O-H repulsion term that is implemented as a bonded interaction[55]. The specifics of bonded and non-bonded interactions have been discussed in detail elsewhere[40,46,56], but it is important to note that all intramolecular nonbonded interactions are excluded by design, which allows for an explicit separation of intra- and intermolecular energies. All simulation cells are constructed from the same nearly orthorhombic supercell which aligns the cartesian z axis as perpendicular to the TATB basal planes and that the shock is along z, constructed using the generalized crystal cutting method (GCCM)[57]. This supercell is replicated from the triclinic P1 crystal structure, with lattice parameters determined by the force field at 300 K and 1 atm, to have supercell vectors A, B, and C, of

$$A = -5a - 3b + 0c$$
$$B = a - 7b + 0c$$
$$C = a + 2b + 6c$$

and had previously been used for a variety of shock studies in TATB[47,58,59]. This cell was the replicated along cartesian axes y and z to form pseudo-2D cells. The z direction was replicated to be approximately equal to lengths of 50, 100, 150, 200, 250, 300, and 400nm. The y direction was replicated to keep an aspect ratio of 2:5. The cell thickness is 4.13nm in all cases. Cylindrical pores were cut perpendicular to the y-z plane, centered at the geometric center of the cell. Cylinder diameters were set to be 1/3 of the cell width, and were roughly 7.2, 14.4, 21.5, 26.3, 33.5, 40.7, and 55.1 nm. Non-periodic boundaries were used in the z direction. Figure 1 shows the initial configuration of the 300nm length (40.7nm diameter pore) and the orientation of the initial supercell.

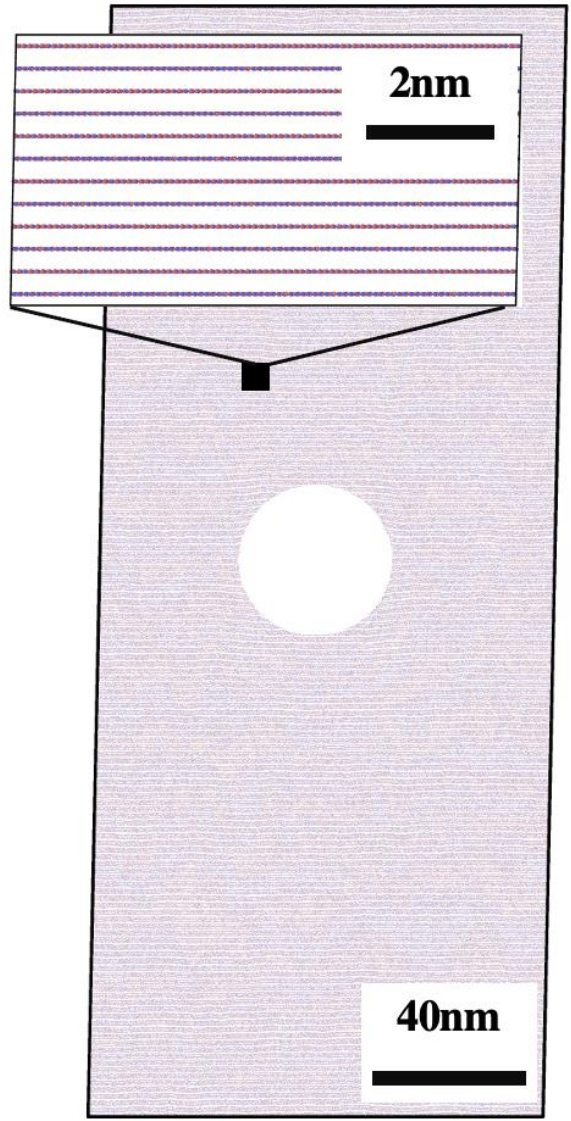

*Figure 1: Initial configuration of 40nm pore system, zoom in shows the orientation of the supercell, with respect to the basal planes of TATB. Supercell has 6 fold rotational symmetry around the vertical axis.*

Prior to impacting, to attenuate the breathing modes associated with creating a free surface in the shock direction, 50 ps of isothermal-isochoric (NVT) dynamics were run, where the first 37.5 ps were broken into 3 sets of 12.5 ps were broken into 10 ps of a Langevin style thermostat[60] and 2.5 ps of a Nose-Hoover thermostat[61]. The last 12.5 ps utilized only a Nose-Hoover thermostat. Shock simulations were run using adiabatic (NVE) conditions. Thermalization was conducted using a 0.5 fs timestep, shock simulations used a 0.2 fs timestep. Shocks were conducted using the reverse ballistic approach[62], setting the bottom 2.5nm of the cell as a rigid piston. Impact velocities used were 0.5, 1.0, 1.5, and 2.0 km/s.

Unless stated otherwise, all analysis performed here was done from a molecularly averaged framework. Positions and velocities were taken to be the center of mass (COM) value for each molecule. The kinetic energy of the molecule was separated into translational ($K_{trans}$) and roto-librational and vibrational ($K_{ro-vib}$) terms. Temperature is taken to be the $K_{ro-vib}$ in units of kelvin by assuming the classical specific heat of $(3N-6)k_b/2$. The intramolecular potential energy ($U_{Intra}$) is defined as

$$U_{Intra} = \sum U_{Bond} + \sum U_{Angle} + \sum U_{Dihedral} + \sum U_{Improper}$$

and the excess, or latent, PE over what is expected from equipartition of energy via the molecule's instantaneous vibrational temperature is defined as

$$U_{Latent} = [U_{Intra} - U_o] - \frac{3N-6}{2} k_b (T - 300)$$

where $U_o$ is the intramolecular potential energy per molecule at 300 K and 1 atm with a perfect, single crystal. All binning was done on the molecular framework in a 2D Eulerian style with bins of 0.7nm by 0.7nm. Normalized temperatures were scaled by a hotspot theoretical maximum temperature for a 1D planar gap[5], defined as

$$\Delta T_{Max} = m U_s U_p / 3 k_b$$

where m is the mass of the molecule. This metric has been previously shown as a good measure for relative hotspot formation efficiency under different conditions[38,40]. Temperature-Area distributions are defined as the cumulative area above temperature T, where local area is defined from the Eulerian binned system.

The relative inwards velocity (RIV) is a normalized metric for how fast the molecules are moving during the collapse process. It is calculated for a given location by taking Eulerian style bins that stretch the diameter of the initial pore area and calculating the net velocity inwards for the particles in that bin. 360 bins are used and atoms are not double counted within bins. Each diameter bin has length of the initial pore and width of 1nm. All bins are centered on the center of the initial pore. Supplemental Material section SM-1 shows a schematic of example bins. Bins are taken for the initial pore area, at some time t. The average of all bins is calculated by weighting each by the number of molecules included in the bin. The relative velocities are used to approximate the relative velocity of all impacts during collapse, such that a particle moving along a diameter at 1.0 km/s, with a particle on the opposite end from the center at rest will have a RIV of 1.0 km/s, whereas two particles moving towards each other at 1.0 km/s and -1.0 km/s from opposite ends of the pore will have a RIV of 2.0 km/s.

## 3. Pore Collapse Mechanisms

Varying the shock strength, as expected, alters the collapse mechanism observed, transitioning from a viscoplastic collapse to a more hydrodynamic mechanism, whereas the change in pore size has limited effect on the transient processes that occur, providing similar collapse mechanisms at the same shock speeds. Figure 2 shows the collapse mechanisms for all sizes and particle velocities performed here, presented in an all-atom framework, colored by particle velocity on a molecule by molecule basis. For $U_p = 0.5$ km/s, the extrusion mechanism occurs for all sizes, with mostly lateral collapse of material in two lobes of plastically worked material expanding into the void space. It should be noted that some collapse motion occurs in the shock direction occurs (note the coloring in Figure 2), but it is slow compared to the lateral event. The lateral event is common for multiple TATB orientations, but the motion in the shock direction varies for different orientations previously studied[13,40]. For $U_p$ = 1.0 km/s, the collapse appears more radial, a mix between the extrusion collapse and the hydrodynamic mechanisms. For $U_p$ = 2.0 km/s and 1.5 km/s, the material undergoes a collapse mechanism that appears mostly as the classical hydrodynamic collapse, where material is focused into a central, accelerating stream moving towards the downstream face of the pore. Nearly all motion is in the direction of the shockwave. However, the material never reaches a true 'hydro', or fluid-like, state, retaining significant crystallinity. Supplemental Material section SM-2 shows an enlarged, hi-res version of the 55.1 nm, 2.0 km/s collapses shown in Figure 2. While the sides of the collapsing lobe of material appear to be amorphous, shear band like material, the central area keeps the basal planes intact. However, these basal planes possess a non-zero defect density. The most central area shows planes that have some buckling defects and non-basal glides, as well as a few small regions of plasticity. The majority of the planes, however, are intact. This region is flanked on either side by areas adjacent to the amorphous regions that are mostly defect free, yet the basal planes have rotated, as material flows, to be almost vertical. The rotation of the planes follows the path pf the shock focusing.

The crystalline nature of the collapsing material and the strength of the hydrogen bonded basal planes can limit how much the material is able to expand and accelerate during collapse, the two key events for high temperature hotspots. Figure 3 shows, for the 2.0 km/s case, the average velocity in the shock direction of the collapsing material (normalized by $U_p$) and the density in the collapsing material (normalized by shock density). Typically, from increasing pore size (hydrodynamic regime), the material is able to expand more before being recompressed on the downstream face of the pore, leading to more P-V work and higher temperatures[5]. Shock focusing, which is related to the curvature of the pore, can lead to higher velocities, which increases both expansion and the re-shock pressure upon impact with the downstream wall, both increasing P-V work and temperature. Figure 3 shows that both of these values reach an asymptotic-like trend for the larger pores, where the velocity sits around $2*U_p$, the expected value for a planar surface[63], and the density does not expand to below the initial density of the material, leveling off at a state that is still compressed with respect with the unshocked state. Previous calculations in HMX with a similar set up resulted in hydrodynamic collapses with velocities near $3*U_p$ and densities well below the initial density where the material behaved quite fluid-like[38].

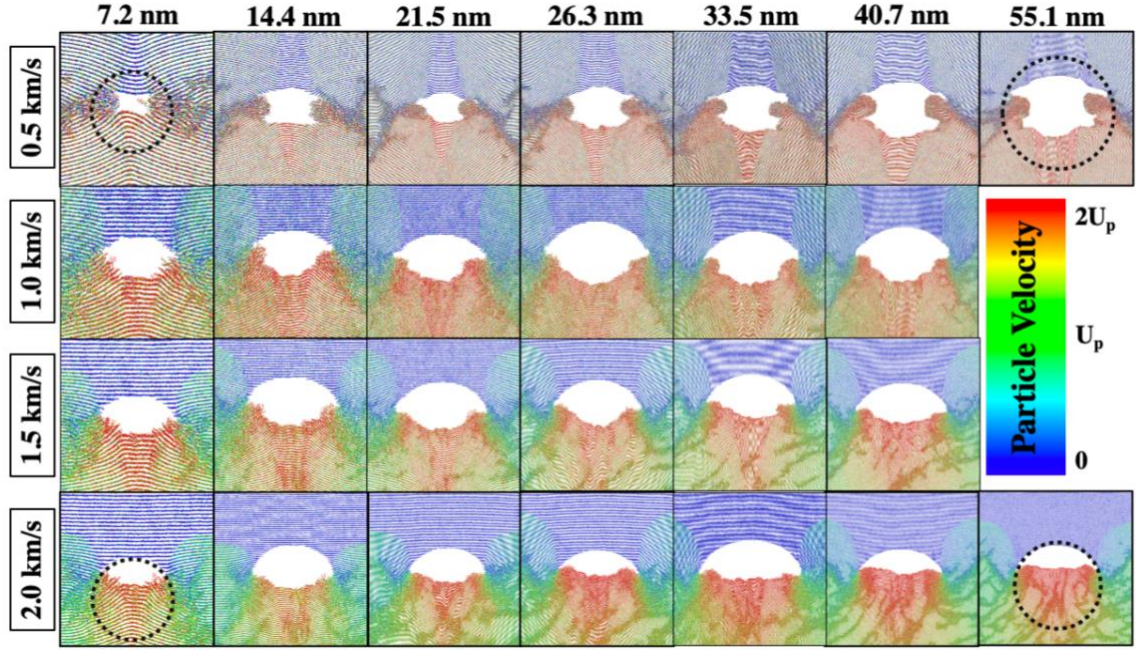

*Figure 2: Collapse processes for all simulations. Color bar is particle velocity (COM velocity in shock direction), from 0 to 2\*$U_p$. Dashed circles represent initial pore area.*

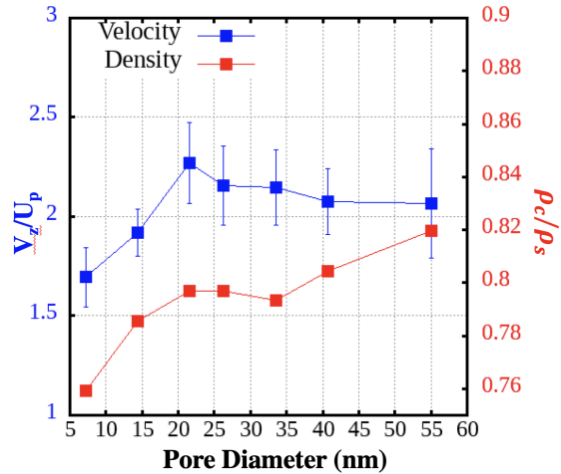

*Figure 3: State of the collapsing material, which is defined as the material in the initial pore area at ~50% collapse by area, for the 2.0 km/s particle velocity shocks. Blue curve and axis are the velocity (in the shock direction) of the collapsing material, normalized by the $U_p$. The red curve is the local density of the collapsing material, normalized by the shock density, 2.66 g/cm³. The value of initial density, 1.85 g/cm³, normalized by shock density is 0.70.*

In Figure 4, the velocities achieved during collapse are analyzed across the various particle velocity cases. For the hydrodynamic cases, the material moves in the shock direction and is easy to measure, however, for weaker shocks, the velocity is mostly lateral. The relative inwards velocity metric, defined in the methods section, calculates (roughly) the average velocity at which material will impact during collapse, independent of which direction it flows. This value, normalized by the initial particle velocity, is presented in Figure 4 for all pore collapse cases studied here. The time of the frame used is determined as the time in which no more than 50% of the initial void area is filled and no material has passed over the center of the initial pore. The hydrodynamic-like collapses, 2.0 and 1.5 km/s, have RIV values of roughly 2 (twice the particle velocity), and the somewhat radial collapse of 1.0 km/s also has a value of roughly 2 for all sizes. For the extrusion collapse, 0.5 km/s, where the two lobes accelerate towards one another, this value reaches upwards of 3. Despite low shock pressure (~5 GPa), the extruded material moves a high velocity and is more plastically worked that the material in the hydrodynamic collapse. Hence, this extrusion collapse mechanism imparts more energy into the collapse for the given shock strength than the stronger shocks that lead to a more radial viscoplastic and hydrodynamic collapse.

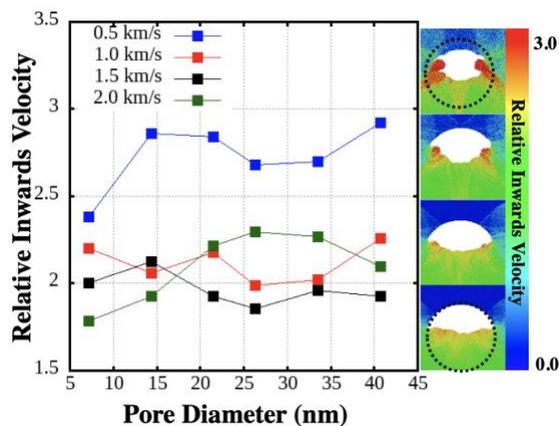

*Figure 4: Plots and heat maps of relative inwards velocity. Scatter plot shows the average RIV for all material inside the initial pore area, as shown by the dotted circles for the 0.5 and 2.0 km/s cases. RIV, which is normalized by the particle velocity, is described in the Methods section. Collapse frames are for 0.5, 1.0, 1.5, and 2.0, from top to bottom, for the 40.7nm diameter case, colored by RIV on a per molecule bases. Dashed circles represent initial pore area.*

## 4. Hotspot Formation
### 4.1. Thermal Energy Localization

To assess the relative, energy localization efficiency of these collapse mechanisms, the resultant hotspots are analyzed. Focus will be placed on the highest and lowest shock speeds in which the collapse mechanisms differ the most from previously studied systems. Figure 5 shows the temperature fields of all hotspots at 5 ps after total volumetric collapse. The color scale is relative to the Up (in meters per second) with upper bounds of 1000, 1500, 2000, and 2500 K for 0.5, 1.0, 1.5, and 2.0 km/s, respectively. As expected, the hydrodynamic collapse appears to create the largest hotspots, and the highest overall temperatures. Hotspots of strong shocks form crescent like shapes, whereas the extrusion mechanism leads to thin hotspots, characteristic of the material flow path during collapse. In all hotspots, the highest temperature regions, denoted by red coloring, are located in areas of impact between collapsing material, either the extruded lobes or the upstream and downstream faces of the pore. Maps of intra-molecular strain energy as available in Supplemental Materials section SM-3.

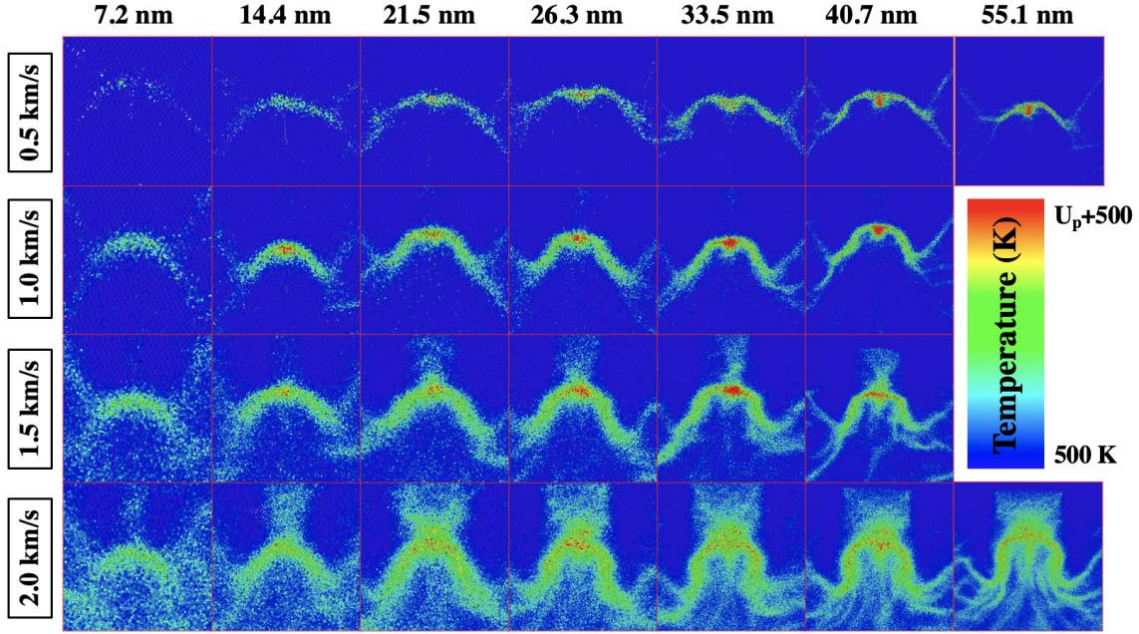

*Figure 5: Color maps of hotspots at time $t_o$ + 5 ps where to is time of complete collapse of porosity. Color bar range is variable for the particle velocity (row by row) where the upper bound is 500 K plus the value of the particle velocity in m/s (1000, 1500, 200, and 2500 K).*

To more quantitatively assess the hotspots, cumulative temperature-area (T-A) distributions are plotted. The T-A plots for all 2.0 km/s shocks are shown in Figure 6, with the left panel showing absolute temperature and area, and the right panel having area normalized by the initial pore size, and temperature normalized by a maximum temperature based on scaling laws for a 1D planar pore[5], $\Delta T_{Max} = mU_sU_p/3k_b$. Most noticeably, the highest temperatures (at low area amounts), which corresponds to the core of the hotspot, reaches an asymptotic value for initial pore areas above 25nm in diameter. This trend is the direct opposite of what has been previously seen for pore size effects in other HE materials[27,38]. This failure to increase temperature for increasing pore size is most likely a direct result of TATB's strong basal planes, which limit the expansion of material during collapse, as shown in Figure 3. Since the material is in the same mechanical state during collapse, despite the different initial pore sizes, the resulting temperature fields are very similar. Additionally, the normalized temperature values do not reach above 0.25, whereas similar void collapses in HMX sat between 0.3 and 0.4 for cylindrical voids 40-80nm.

The lack of increased efficiency and higher temperatures for larger void sizes may contribute to TATB's insensitivity compared to more traditional HE materials in which larger pores are able to support additional material expansion and recompression, leading to temperatures much higher than those found here.

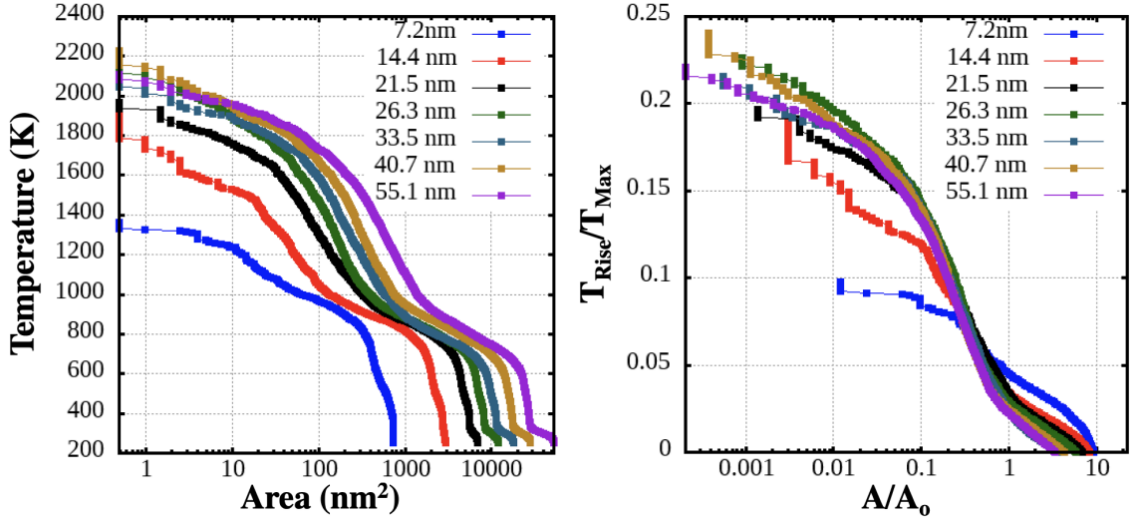

*Figure 6: Temperature-Area distributions, as described in Methods, for all 2.0 km/s particle velocity pore collapses, at time $t_o + 5$ ps, where $t_o$ is time of complete collapse of porosity. The right hand plot is temperature above the shock temperature, normalized by the 1D hotspot maximum temperature scaling law from Ref. [5].*

Figure 7 shows the same style T-A plots as in Figure 6, but for the 0.5 km/s cases, which exemplify the extrusion mechanism of collapses. Previous works have shown viscoplastic dominated collapses (radial) to be slightly more efficient at localizing temperature than the hydrodynamic case, with 0.5 km/s cases in 80nm pores of HMX reaching between 600-700 K at peak temperatures, with normalized values of ~0.5[38]. Here, the extrusion collapses in TATB reaches surprisingly higher temperatures, over 1000 K peak temperatures for the largest pores, with normalized temperatures near 1.0. For HMX, strong shocks with high aspect ratio voids that cause molecular ejecta (gas density) were required to reach such normalized values of temperature[38]. The high temperatures and efficiency of these hotspots are directly related to the large RIV values during collapse (see Figure 3). The extruded lobes move towards each other at speeds over 1.5 km/s each, leading to impacts at relative velocities almost as high as the hydrodynamic collapse, which reaches 4 km/s before impacting off of the stationary downstream face. These high velocities reached in the extrusion create much higher temperature values than could be expected from a more typical cavitation like event at similar pressures.

When normalized by the expected maximum temperature values for low shock speeds, these temperatures give extremely high normalized values. It should be noted that the hotspots from the extrusion mechanism are significantly smaller than the hydrodynamic collapse, and may not contribute significantly to material ignition. These extrusion formed hotspots are of similar size and excess temperatures to a single shear band[56]. Additionally, the continued growth of the temperature value for increasing pore size differs from the trend shown in the hydrodynamic collapse, which stagnates at larger sizes. However, collapse mechanisms tend towards a hydrodynamic nature at low pressures for larger pores[26], and it is unknown whether the extrusion mechanism will occur at micrometer (or larger) sized pores where the hotspot size would be much more relevant to chemical ignition.

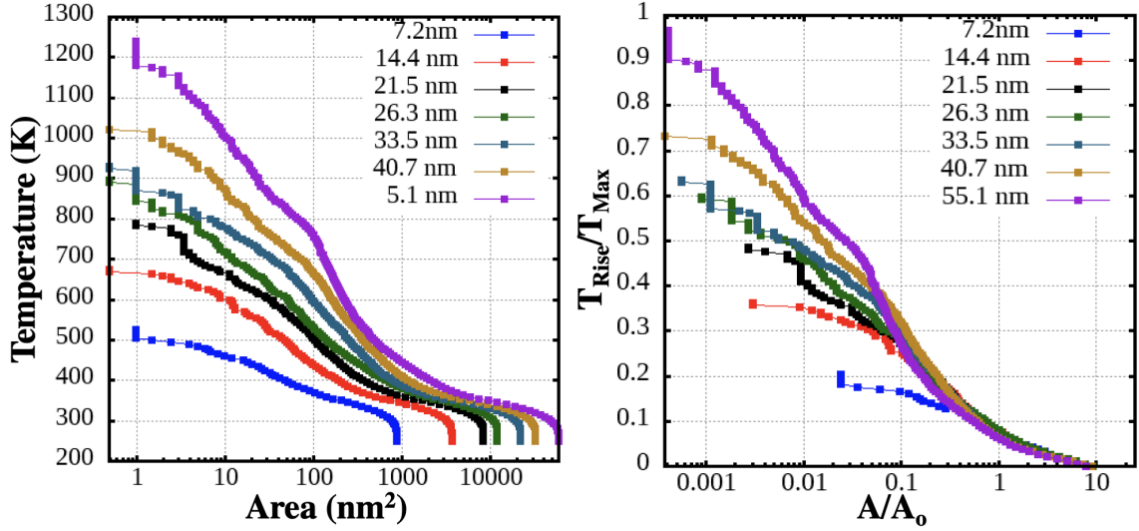

*Figure 7: Temperature-Area distributions, as described in Methods, for all 0.5 km/s particle velocity pore collapses, at time $t_o + 5$ ps, where $t_o$ is time of complete collapse of porosity. The right hand plot is temperature above the shock temperature, normalized by the 1D hotspot maximum temperature scaling law from Ref. [5].*

Figure 8 shows the T-A plot trends for increasing shock strength for the 40.7 nm diameter pore. As expected, the absolute temperature increases with increasing shock strength. The opposite trend exists for the normalized temperature, where, interestingly, the range of values is from 0.2 to 0.75. This is driven in part by the extrusion mechanisms of weak shocks providing much more thermal energy than a typically viscoplastic collapse, and in part by the hydrodynamic collapse being limited due to the strongly bonded basal planes remaining intact during collapse, limiting the P-V work done during recompression. For comparison, this range of normalized values for HMX 40nm cylindrical pore collapse is roughly 0.3-0.5[38].

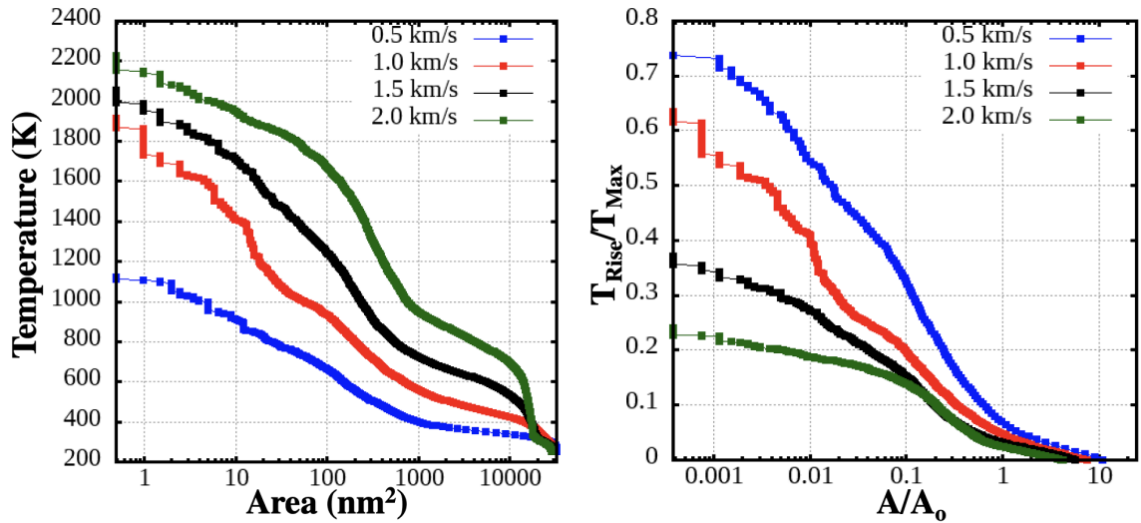

Figure 8: *Temperature-Area distributions, as described in Methods, for all 40.7 nm pore collapses, at time $t_o$ + 5 ps, where $t_o$ is time of complete collapse of porosity. The right hand plot is temperature above the shock temperature, normalized by the 1D hotspot maximum temperature scaling law from Ref [5].*

### 4.2. Intra-Molecular Strain Localization

While the thermal energy localization in the hotspot is the most important factor for prompt chemistry, the intra-molecular strain energy produced by shock-driven, shape perturbations in the molecule can have a considerable effect on local kinetics and reaction pathways[47]. These intra-molecular strains do not track with the temperature on a molecule by molecule basis and are most likely governed by different mechanisms[46]. Figure 9 shows the cumulative intra-molecular strain energy - area distributions for the 2.0 and 0.5 km/s shocks. This 'latent potential energy' is the excess potential energy on top of the thermal contributions and is driven mainly by molecular strain. As expected, the stronger shocks lead to significantly more molecular strain. Interestingly, in the opposite of temperature, the weak shocks seem to have little trend with the pore size, whereas the stronger shocks have an increasing energy with increasing pore size.

Previous work on intra-molecular strain energy[40], as well as the maps of it shown in Supplemental Material section SM-3, show that it manifests where the most plastic flow occurs. For the extrusion mechanisms of weak shocks, this localizes in the core of the hotspot where the extruded material ends up. For the hydrodynamic collapse, this is, in addition to the core, the area upstream of the peak temperatures, where significant volume must flow to fill in the pore. This is the material that is highly worked from the shock focusing.

It follows that more plastic flow is needed for larger pores when in the hydrodynamic regime, and the rapid collapse leads to more strain energy that cannot relax due to the high density of the hotspot. The extruded material at low shock speeds, which flows more readily to fill the void and has a much lower shock density, may not recompress fast enough or to high enough density to deform the molecules into strain states that do not relax to the ground state quickly. Figure 10 shows the distributions for increasing shock strength, which gives the same trend as temperature, increasing with increasing shock strength. This corroborates the claim that plastic flow is what leads to excess potential energy, which can be expanded to state that *rapid* material flow that leads to a high density state is the necessary recipe for large intra-molecular strains that lead to mechanochemical acceleration of chemistry. This tracks well with the idea that shear band formation from strong shocks leads to large intra-molecular strains[56].

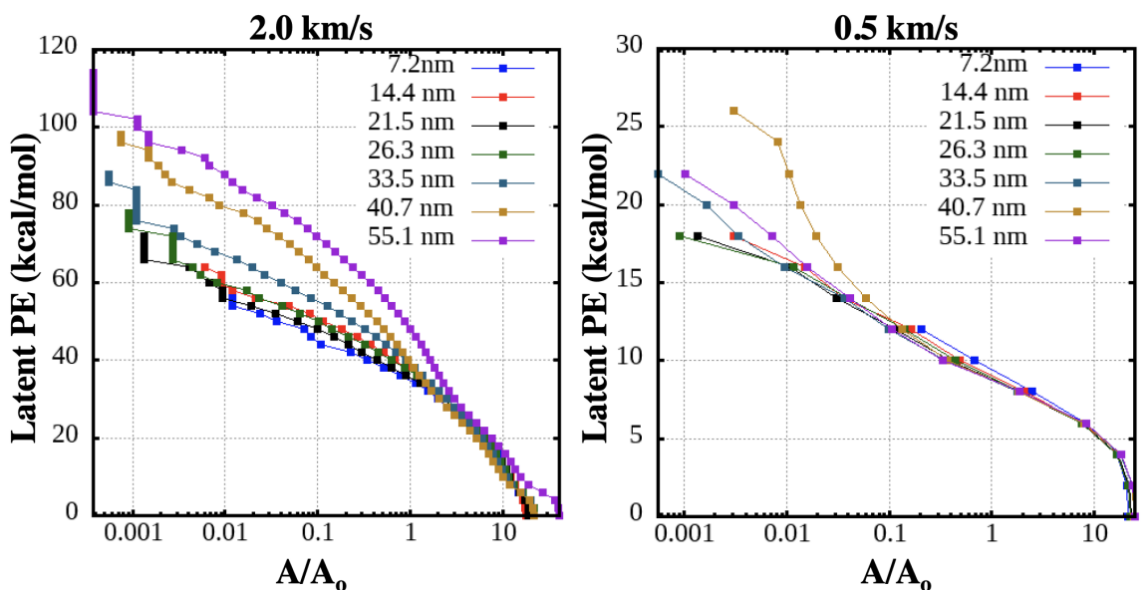

*Figure 9: Area normalized distributions of intra-molecular strain energy (or 'Latent PE') for all 2.0 km/s and 0.5 km/s pore collapses.*

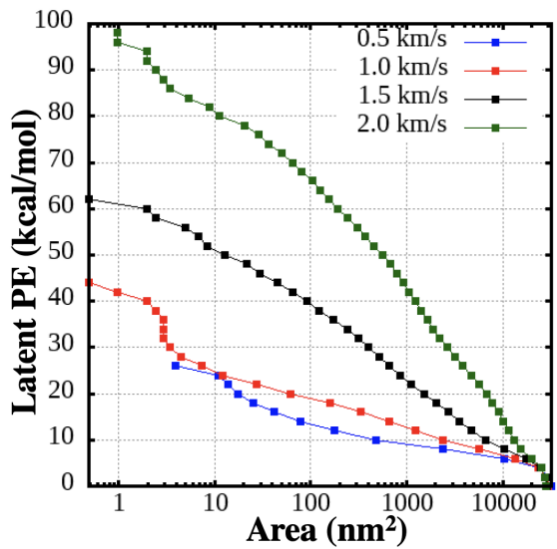

*Figure 10: Area normalized distributions of intra-molecular strain energy (or 'Latent PE') for all 40.7 nm pore collapses.*

# 5. Conclusions

All-atom MD simulations were conducted to assess the shock induced pore collapse in TATB cylindrical pores, resulting in complex collapse mechanisms atypical of other molecular crystals. At high shock speeds, the collapse occurred with the shock, pushing the upstream face of the pore into the void space. However, this material does not go to a hydrodynamic state, keeping the strong, hydrogen bonded basal planes intact, which prevents the material from expanding significantly. This results in a lack of high temperature hotspots, and increasing the pore volume does not result in hotter hotspots, as has been found with other systems in the past. For weak shocks, the viscoplastic response is an extrusion like mechanism, with plastically worked lopes ejecting laterally from the sides of the pore. This results in high velocity impact of the collapsing material and temperatures nearly double what is found for traditional, radial collapses. However, these hotspots are significantly smaller in area than the strong shocks. The intra-molecular strain energy (a latent potential energy) was also assessed for these collapse mechanisms, showing that this localization, which does not track with temperature, tracks with areas of material flow. This helps to show that this localization of energy, which leads to a mechanochemical acceleration of kinetics, is nucleated through a rapid flow of material that results in a high pressure state which prevents significant intra-molecular relaxations. These results provide a significant advancement of the understanding of hotspot formation mechanisms from pore collapse, especially for strongly bonded systems such as TATB where strength of the material still matters in the "hydrodynamic" regime and the viscoplastic response leads to complex collapse mechanisms. Additionally, this reveals a direct mechanistic recipe for localizing intra-molecular strain.

# Acknowledgements

Funding for this project was provided by the Director's Postdoctoral Fellowship program, project LDRD 20220705PRD1. Partial funding was provided by the Advanced Simulation and Computing Physics and Engineering Models project (ASC-PEM). This research used resources provided by the Los Alamos National Laboratory Institutional Computing Program. This work was supported by the U.S. Department of Energy (DOE) through the Los Alamos National Laboratory. The Los Alamos National Laboratory is operated by Triad National Security, LLC, for the National Nuclear Security Administration of the U.S. Department of Energy (Contract No. 89233218CNA000001). Approved for unlimited release: LA-UR -22-31052.

# Supplemental Material to:
# Energy Localization Efficiency in TATB Pore Collapse Mechanisms

Brenden W. Hamilton*, Timothy C. Germann

Theoretical Division, Los Alamos National Laboratory, Los Alamos, New Mexico 87545, USA

* brenden@lanl.gov

# SM-1

For the Relative Inwards Velocity (RIV) calculations, Eulerian bins aligned with pore diameters were used. Figure SM-1 below shows examples of these bins for the 0.5 km/s case.

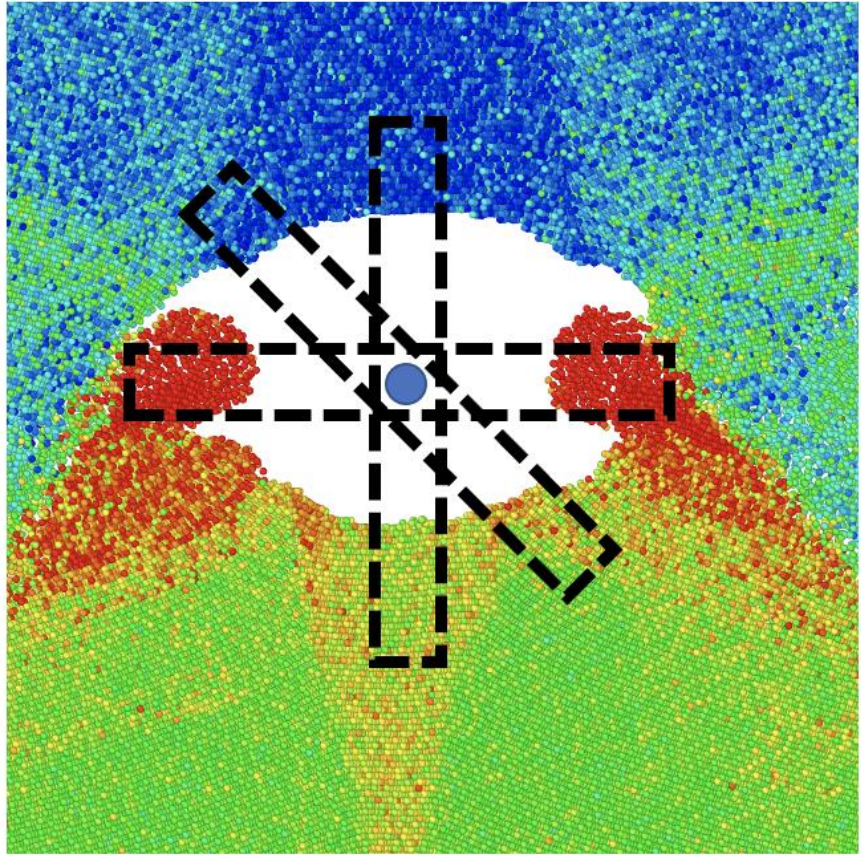

*Figure SM 1: Diameter Bins for 40.7nm pore shocked at 0.5 km/s. Molecules are colored by RIV.*

SM-2

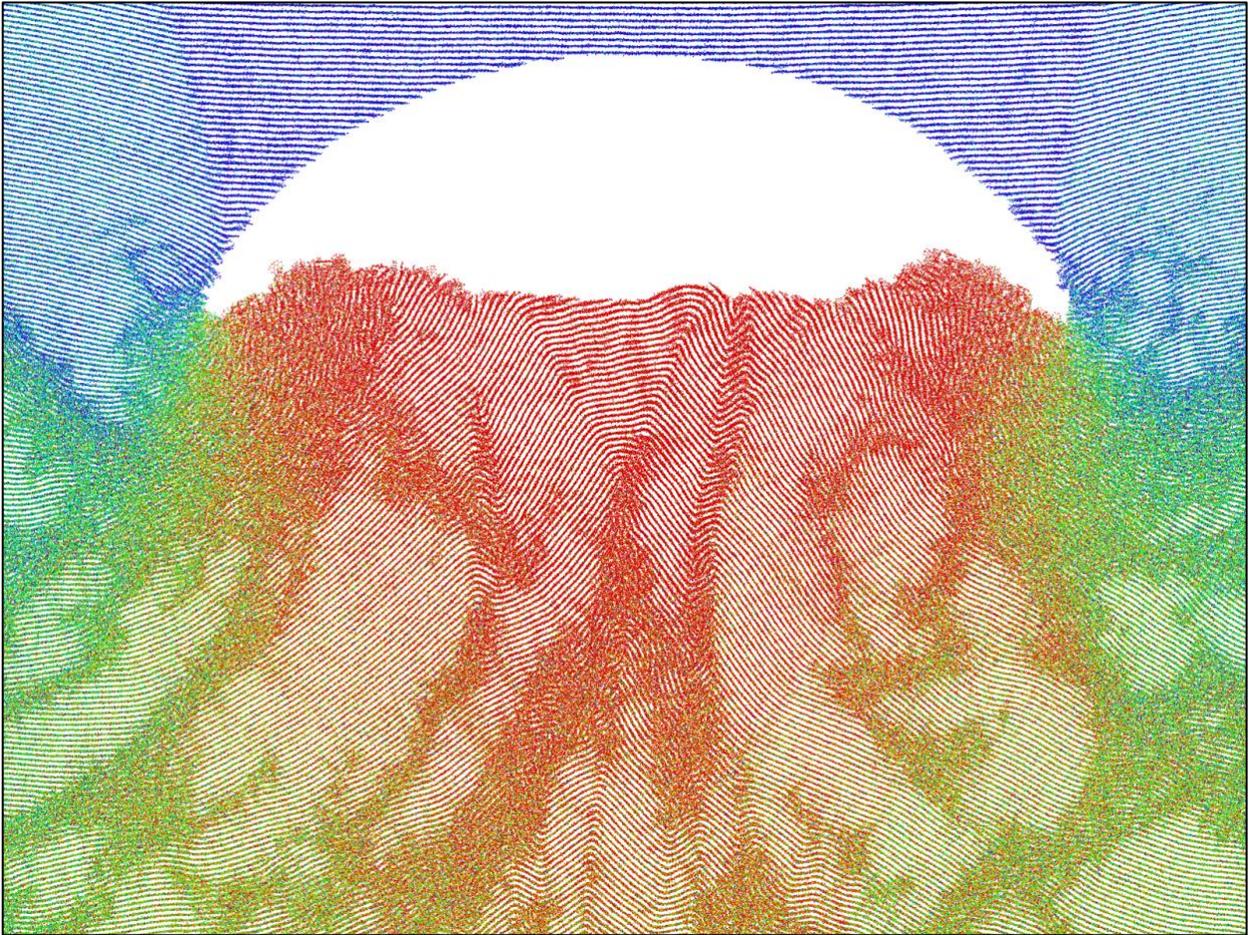

*Figure SM 2: High resolution image of the pore collapse for the 55nm case at 2.0 km/s. All atoms are rendered and colored by velocity in the shock direction.*

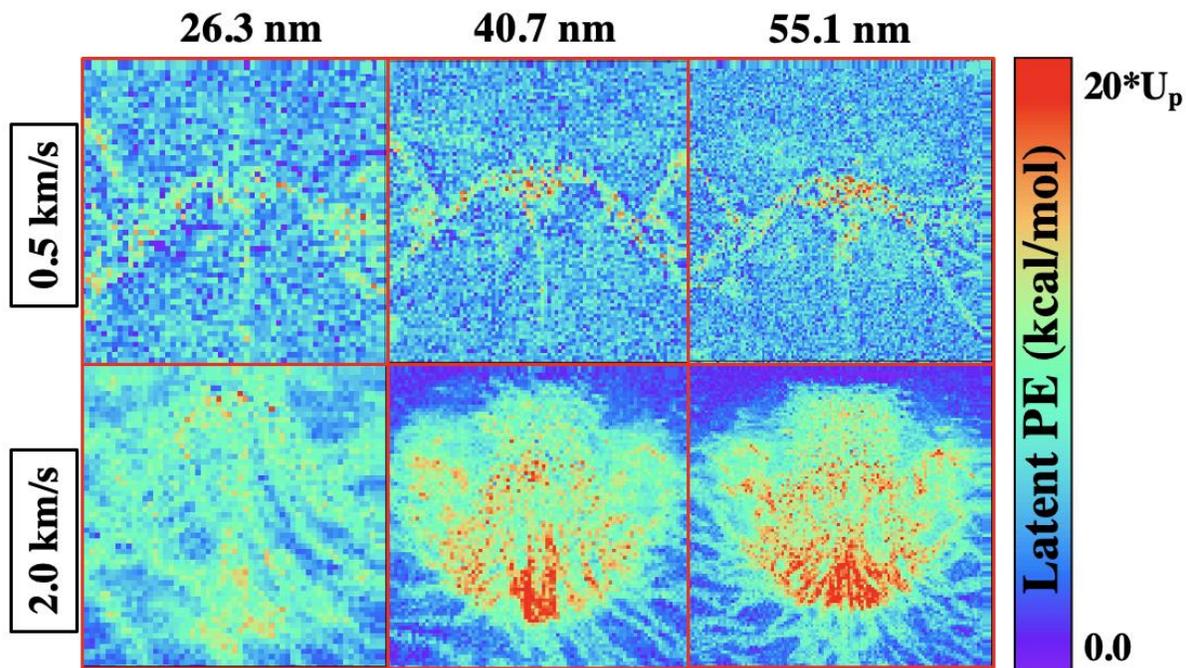

*Figure SM 3: Eulerian bin maps of latent potential energy, the intra-molecular strain energy, for 0.5 and 2.0 km/s cases for 3 different pore sizes. Image scales are set such that the rendering lengths are 1.5D * 1.5D where D is the pore diameter.*

## Acknowledgements

Funding for this project was provided by the Director's Postdoctoral Fellowship program, project LDRD 20220705PRD1. Partial funding was provided by the Advanced Simulation and Computing Physics and Engineering Models project (ASC-PEM). This research used resources provided by the Los Alamos National Laboratory Institutional Computing Program. This work was supported by the U.S. Department of Energy (DOE) through the Los Alamos National Laboratory. The Los Alamos National Laboratory is operated by Triad National Security, LLC, for the National Nuclear Security Administration of the U.S. Department of Energy (Contract No. 89233218CNA000001). Approved for unlimited release: LA-UR -22-31052.

[63] J.W. Forbes, *Shock Wave Compression of Condensed Matter* (Springer Science & Business Media, 2013).

21